\documentclass[12pt,a4]{article}
\usepackage{mathtext}
\usepackage{graphicx}
\usepackage{amssymb,amsfonts,amsmath}
\usepackage[small]{caption2}
\usepackage{epsf}

\usepackage{rotate}
\parskip=\baselineskip	

\makeatletter

\renewcommand{\section}{\@startsection
	{section}{1}%
        {0mm}%
        {1.5\baselineskip}%
        {0.5\baselineskip}%
        {\centering\normalfont\normalsize}}%

\renewcommand{\subsection}{\@startsection
	{subsection}{2}%
        {0mm}%
        {-\baselineskip}%
        {\baselineskip}%
        {\centering\itshape\normalsize}}%

\renewcommand{\@biblabel}[1]{#1.}

\makeatother
\newcommand{\Section}[1]{\section{\uppercase{#1}}}

\begin{document}


\begin{center}
{\large\bf\uppercase{Magnetic fileds of coalescing neutron stars and
the luminosity function of short gamma-ray bursts\\}}
{\large\bf
K.\,A.\,Postnov, A.\,G.\,Kuranov
}
\end{center}

{\it
\leftskip=1cm
\parindent=-\leftskip
\parskip=0pt
Sternberg Astronomical Institute, 119992, Moscow, Russia

}


\begin{abstract}

\footnotesize
Coalescing neutron star binaries are believed to be the most 
reliable sources for ground-based detectors of gravitational waves and 
likely progenitors of short gamma-ray bursts. 
In the process of coalescence, magnetic fields of neutron stars
can induce interesting observational manifestations and affect 
the form of gravitational wave signal. In this papaer we use 
the population synthesis method to model the expected  distribution of neutron 
star magnetic fields during the coalescence under different assumptions 
on the initial parameters of neutron stars and their magnetic field evolution. 
We discuss possible elecotrmagnetic phenomena preceding the 
coalescence of magnetized neutron star binaries and the effect of
magnetic field on the gravitational wave signal. We find that 
a log-normal (Gaussian in logarithms) distribution of the initial 
magnetic fields of neutron stars, which agrees with observed properties
of radio pulsars, produces the distribution of the magnetic field energy during
the coalescence that adequately describes the observed luminosity function of 
short gamma-ray bursts under different assumptions on the field evolution 
and initial parameters of neutron stars. This agreement lends further support 
to the model of coalescing neutron star  
binaries as progenitors of gamma-ray bursts.

\end{abstract}


\section{Introduction}

Neutron stars (NS) are end products of thermonuclear evolution of 
stellar cores with a mass above  $\sim 8-10 M_\odot$. The birth of a NS 
in the core collapse is accompanied by the ejection of the outer stellar
envelope which is observed as type II or Ib/c supernova, depending on the evolutionalry 
status of the star immediately before the collapse. Young NS observed as
radio pulsars demonstrate large space velocities (typically several hundreds
km/s, see e.g. Hobbs et al. 2005), which is commonly explained by a hypothetical
kick velocity acquired by a nascent NS during the collapse. Different physical mechanisms
of the kicks are duiscussed in the literature (e.g. Lai 2001). 

It is well known that astrophysical manifestations of a NS at all
evolutionary stages are determined by its spin, magnetic field, and 
the interaction with the of the surrounding medium (see Lipunov 1992). 
The evolution of a NS in a binary system is even more complicated
by the presence of the secondary component.  So the properties of 
the galactic population of NS in binaries are subjected to  
uncertainties in the theory of evolution of binary stars, such as 
the initial orbital parameters of binary stars, uncertainties in the description 
of the evolutionary mass loss of normal stars via stellar winds, 
the efficiency of common envelops, etc. Binary NS observed as a specific 
subclass of binary pulsars are of special interest. Close binary NS (also 
called relativistic double stars) are considered as best sources 
of gravitationla waves to ve observed by the present-day gravitational 
wave detectors (Grishchuk et al. 2001). The evolution of relativisitc binaries 
and binary pulsars is considereed in more detail in many reviews 
(see, e.g., Postnov and Yungelson (2006), Bisnovaty-Kogan (2006)).  
The coalescence of binary NS due to secular decrease of the orbit
because of orbital angular momentum removal by gravitational waves 
can be the primary formation channel of short hard gamma-ray bursts
(Blinnikov et al. 1984, Eichler et al. 1989, see Nakar 2007 for 
extensive review).  

Magnetic fields of coalescing NS binaries can lead to intersting observational manifestations preceding the gravitational wave burst (e.g. Lipunov and Panchenko 1996, 
Lipunova et al 1997), and can play a significant role in shaping 
the expected form of gravitational wave signal (Lipunova and Lipunov 1998). 
The recent general relativistic numerical calculations of coalescence of magnetized  double NS (Yuk Tun Li et al. 2008) cofirmed the crucial role of strong magnetic 
field in delaying the collapse of an intermediate super-massive object 
that forms during the coalescence process. Obviously, the magnetic field
cannot be neglected when calculating the gravitational-wave templates from
colaescing NS binaries, which is very important for ongoing gravitational wave detectors
(Braginsky 2008, Abbott et al. 2008).

The population synthesis method is a powerful tool to study astrophysical 
manifestations of NSs. It enables calculation of large ensembles of objects 
and testing different scenarios of their evolution (see, e.g., Popov and
Prokhorov 2007 for more detail). This method, with different degree of 
account for features of binary star evolution, has been successfully applied to study 
the evolution of coalescing NS binaries by differnt authors (Tutukov and Yungleson 1994, 
Lipunov et al. 1997, Portegies Zwart and Yungelson 1998, Belczynski et al. 2008). 
In recent papers (Postnov and Kuranov 2008, Kuranov et al 2009), the 
population synthesis method was applied to study the effect of the possible 
kick-spin correlation on the coalescence rate of binary NSs and the
observed correlation of the spin axis of radio pulsars with the direction of
their space velocities. 

In the present work we use the Scenario Mashine code (see
the detailed description in Lipunov et al. 2009) to calculate the 
expected distribution of magnetic fields in coalescing binary NSs 
under different assumptions on the initial distribution of NS magnetic field
and its evolution (no field decay or accretion-induced decay). The calculated 
distribution of magnetic fields is comapared with the observed 
luminosity function of short gamma-ray bursts. It 
is shown that for the initial log-normal distribution of NS magnetic fields, 
which well describes observational poroperties of radio pulsars,   the calculated 
distribution of the magnetic energy (both the 
energy of the poloidal magnetic field and the total magnetic field including the toroidal component) in coalescing binaries well fits the observed luminosity function of 
of short gamma-ray bursts if one assumes that the total energy release in a gamma-ray burst formed during the coalescence is proportional to the total magnetic field energy. 
The agreement is achieved in different models of the NS magnetic field evolution
(the field decay time, accreiton-induced decay) and initial p[arameters of young NS
(the amplitude of natal kick velocity and its possible correlation with the NS spin).

\Section{The model}

\subsection{Magnetic fields of neutron stars}

The initial magnetic field of neutron stars is one of the main parameters of our calculations. We used two possible forms of the initial magnetic field distribution
of NSs: 1) log-normal distrubution of dipole magnetic moments
$\mu$:
\begin{equation}
  f(\mu)\,d\mu \propto
  	\exp\left[-\frac{(\log\mu-\log\mu_0)^2}{\sigma_\mu^2}\right]d\mu\,,
\label{mu-norm}
\end{equation}
and 2) plain in logarithm distribution: 
\begin{equation}
  f(\mu)\,d\mu \propto
  	\frac{\log\mu}{\log\mu_{\max}-\log\mu_{\min}}d\mu\,.
\label{mu-flat}
\end{equation}

The first law directly follows from the analysis of 
the observed properties of radio pulsars and the field estimation
using pulsar spin period and its first derivative 
(see the ATNF pulsar catalogue, Hobbs and Manchester 2009). 
The second law covers a broad range of magnetic field values 
and assumes an equal birth probability for strong-field NSs (like magnetars) and 
low-field NSs (like anti-magnetars, i.e. radio pulsars with low
surface fields and slow rotation). Indeed, some recent studies (e.g. Woods 2008) 
suggest the birthrate of magnetars comparable with that of ordinary pulsars, 
and one of the youngest NSs in the center of the supernova remnant Cas A 
is likely to be an anti-magnetar (Gotthelf and Halpern 2008). 
Of course, the NS magnetic filed at birth can depend on the NS parameters
(for example, on its angular momentum), but in view of the existing 
uncertainty in NS formation theory we do not consider 
such a complication of the initial conditions.

In our calculations we adopted the following parameters of the initial 
distribution of the NS dipole magnetic moment (Faucher-Giguere and Kaspi 2006):
\begin{eqnarray}
  \log\mu_0=30.35& , & \sigma_{\mu}=0.55,\\     
  \mu_{\min}=10^{26}~\textrm{Гс\,см}^3 & & (B_{\min}=10^{8}~\textrm{Гс}),\nonumber \\
 \mu_{\max}=10^{32}~\textrm{Гс\,см}^3 & &(B_{\max}=10^{14}~\textrm{Гс}), \nonumber
\end{eqnarray}
(the NS surface magnetic field 
$B$ is calculated for the assumed NS radius $R_{\rm NS}=10$~km).

\subsection{Magnetic field evolution}

The evolution of the NS magnetic field is poorly known. It follows from 
general theoretical considerations that poloidal magnetic fields of single NSs must 
decay due to Ohmic losses in the crust (Flowers and Ruderman, 1977)
with the characteristic time longer than $\sim 10^7$~yrs. Observational 
tests of this hypothesis yield controversial results. For example, 
the analysis of statistical properties of radio pulsars leads some authors to conclude that the magnetic field of NSs must decay with the characteristic time 
 $\sim 3$~mln yrs (Gontier et al 2004), while another authors argue 
that statistical properties of radio pulsars do not require any magnetic field decay
on timescales as long as comparable to the pulsar life time
 $\sim 100$~mln yrs (Faucher-Giguere and Kaspi 2006). 

In our calculations we used the following parametrization of the NS 
magnetic field evolution with time:
\begin{equation}
\mu=  \begin{cases}
\mu_0 e^{-t/\tau}, \quad  \mu > \mu_{\min},\\
               \mu_{\min}, \quad \quad      t > \tau \ln(\mu_0/\mu_{\min}). 
	\end{cases}	
\label{mu(t)}        
\end{equation}
The characterisitc decay time of the dipole magnetic field 
was chosen from $\tau=10^7$~yrs to $\infty$ (no decaY), 
and the minimum possible NS magnetic dipole moment was taken to be 
$\mu_{\min}=10^{26}$~G$\cdot$сcm$^3$, which corresponds to a NS surface
field of $B\simeq10^8$~G.

Indeed, observations and theoretical arguments suggest 
that the surface poloidal magnetic field of an accreting NS in a close binary system
must asymptotically decay (Bisnovaty-Kogan and Komberg 1974, 
Taam and Van den Heuvel 1986, Romani 1990, Bhattacharia 2002) 
down to a lower limit of  $10^8-10^9$~G (Kulkarni 1986, Van den Heuvel et al. 1986).
The magnetic field decay in accreting NSs was accounted for in the form (Romani 1990):
\begin{equation}
	\mu = \mu_0 (1+10^6\Delta m)^{-0.8}\,,
\label{accr-decay}
\end{equation}
where $\mu_0$ and $\mu$ are the value of the NS dipole magnetic moment
before and after the accretion stage, respectively, $\Delta m$ is the mass of 
accreted matter (in solar masses). 

\subsection{Toroidal field of NSs}

The stability analysis of an axisymmetric magnetic field 
(Prenderghast 1956, Tayler 1973, Wright 1973) suggests that a substantial fraction of 
the total magnetic field energy of a NS is contained in the toroidal component. 
Each field component (poloidal or toroidal) are separately unstable but can be
stabilized by the presence of another component. The stability criterion of
an axisymmetric magnetic field depends on several parameters and for a NS can be 
expressed in terms of the ratio of the energy of the poloidal component  
$E_p$ to the total magnetic energy of the star $E_m=E_p+E_t$ 
(see Braithwaite 2008):
the upper limit for stability is $(E_p/E_m)\sim 0.8$;
the lower limit for stability is 
$(E_p/E_m)_{crit}\sim 10^3E_m/|E_{gr}|$.
Here $E_{gr}$ is the gravitational binding energy ($\sim 10^{53}$ erg). 
In our calculations we assumed that the NS magnetic field always
satisfies this stability criterion, i.e. for a given value of 
$E_p(t)$ the total magnetic field energy $E_m(t)$ falls within the 
stability limits
\begin{equation}
1.25E_p(t) < E_m(t) < 10^{-3}\sqrt{E_p(t)|E_{gr}|}.
\label{stability}
\end{equation}

\subsection{Kick velocity during NS or BH formation}

The collapse asymmetry leads to additional kick velocity to 
a newborn NS. We modeled this velocity as follows: 
the direction of the kick velocity vector was taken to lie randomly within
a cone with angle 
$\theta$ around the rotational axis of the pre-supernova, and its absolute
value was assumed to be distributed according to a Maxwellian law:
\begin{equation}
	f(v)\,dv \propto \exp\left(-\frac{v^2}{V_m^2}\right)\,v^2\,dv,
\label{kick-maxwell}
\end{equation}
where $V_m$ is the characteristic velocity. 

We have used two frequently considered kick velocity distributions
(see the discussion in review Postnov and Yungelson (2006) and in the recent
paper Martin et al. 2009).

\textbf{Model А}. Single-mode Maxwellian distribution.
All NSs acquire kick velocity at birth. The parameter  
$V_m$ varies within the range $100-500$~km/s.

\textbf{Model В}. Bimodal Maxwellian distribution. For the components of
close binary system (here we call a system as close binary 
if one of the components filled its Roche lobe and mass transfer occurred at 
some stages preceding the core collapse supernova of the primary component) 
with initial masses $8M_\odot$--$11 M_\odot$, the electron-capture supernova
from degenerate O-Ne-Mg stellar core is assumed. In that case the natal kick is
small or zero. This assumptions allows one to explain some observational properties 
of X-ray Be-systems and binary pulsars 
(see Podsiadlowski et al 2004, Van den Heuvel 2004, 2007). 
In Model B with bimodal kick we assumed  
 $V_m=30$ км/с for a pre-supernova with the initial mass from the range 
 $8M_\odot$--$11 M_\odot$ and the mass transfer stage took place before the collapse. 
In other cases (i.e. for NS born from single progenitors or binary components with 
the initial mass above 11 solar masses) the parameter  
 $V_m$ varied within the range $100-500$~km/s, as in Model A. So 
\[ V_m = \left\{
\begin{array}{rll}
100 ... 500\mbox{km/s}, & > 10M_\odot,& \mbox{Model A}\\
\\
100 ... 500\mbox{km/s}, & > 11M_\odot,& \mbox{Model B}\\
30\mbox{km/s},  &   8M_\odot$--$11 M_\odot, close binary
\end{array}\right.\]

For both kick models we assumed all single NSs and NSs in wide binaries to have
the initial mass $M_{\rm NS}=1.4M_\odot$. In model B, NSs formed from e-capture SN
are assumed to have smaller initial masses $M_{\rm NS}=1.25M_\odot$.

\Section{Results of population synthesis calculations}

The models of the natal kick and magnetic field evolution of netron stars are 
listed in Table 1. The results of calculations of different models are
presented in Fig. 1-5. For each model, calculations were performed for 
isotropic kicks (corresponding to   
$\theta=90^{\circ}$) and kicks collimated with the 
rotational axis of the NS within the cone $\theta=10^{\circ}$. This value of
$\theta$ is suggested by the recent analysis of the spin-velocity alignment 
of radio pulsars (Kuranov et al. 2009). 

\begin{figure*}
\centering
\includegraphics[height=1.0\textwidth, width=1.0\textwidth]{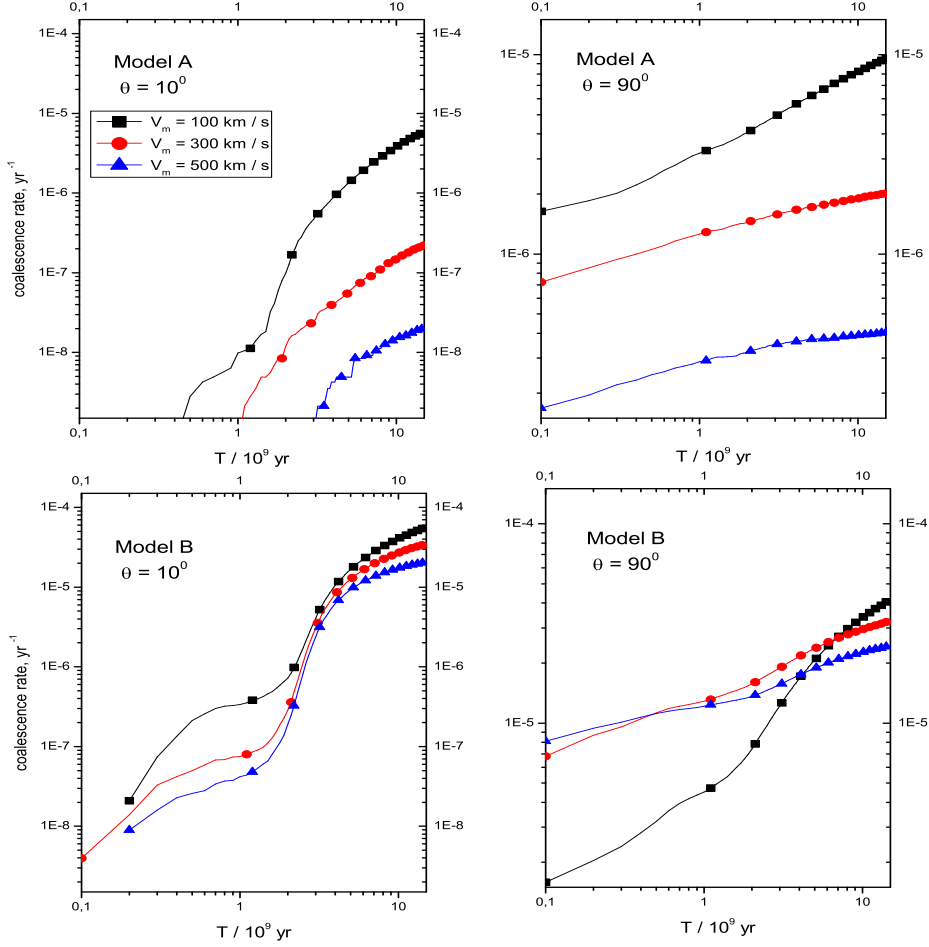}
\caption{Integral distribution function of the binary NS coalescence rate 
as a function of time between the birth and merging for different NS formation 
parameters. The rate is normalized to the galactic star formation 
rate $3M_\odot$ лет$^{-1}$. Upper plots: model A, bottom plots: model B. 
Plots to the left: the natal kick-spin alignment within the cone with angle 
$\theta=10^{\circ}$. Plots to the right: the isotropical natal kick
($\theta=90^{\circ}$).
\label{t_cum_lgA}}
\end{figure*}

\begin{figure*}[ht]
\centering
\includegraphics[height=0.6\textwidth,width=0.6\textwidth]{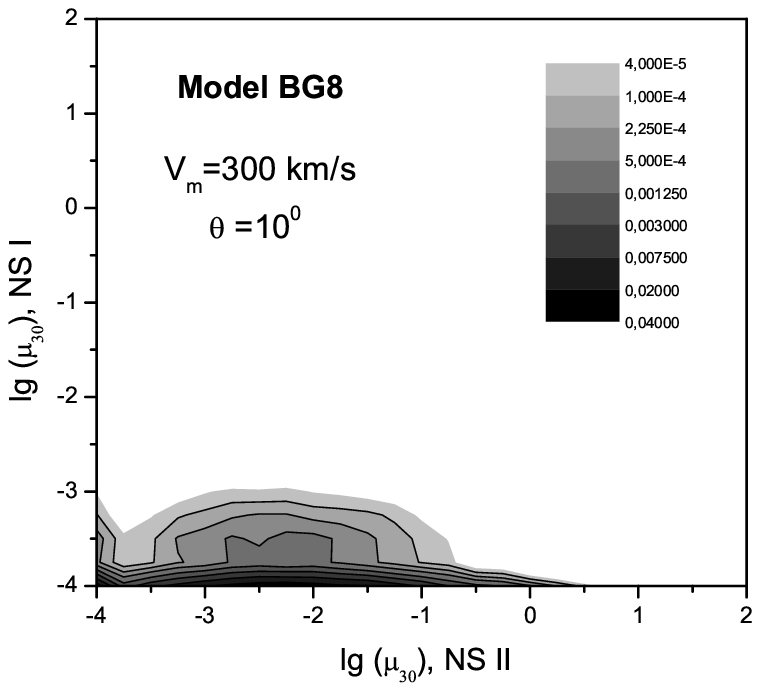}
\includegraphics[height=0.6\textwidth,width=0.6\textwidth]{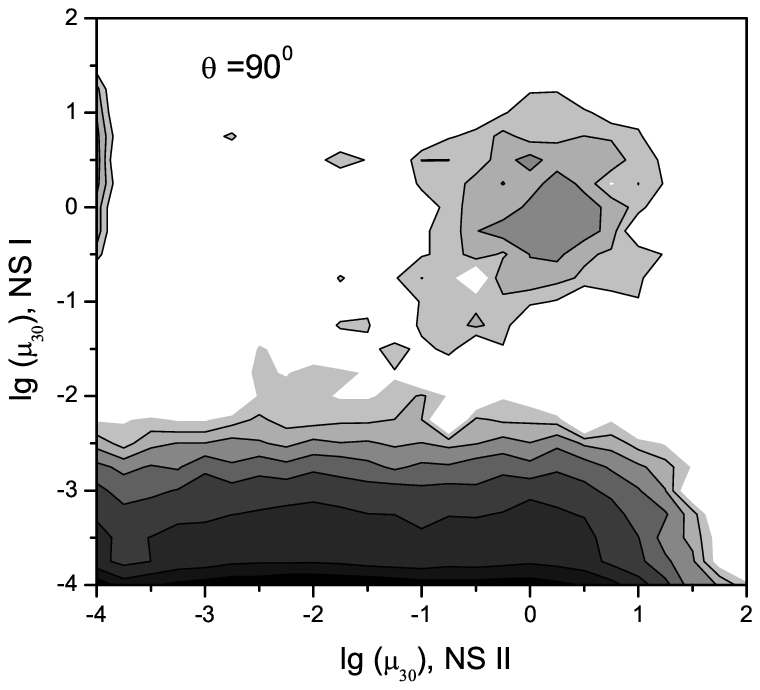}
\caption{The distribution function of dipole magnetic 
moments of coalescing meutron stars 
(NSI denotes the first (older) NS, NS II -- the second (younger) NS in the binary)
by the moment of coalescence in Model BG8. 
The upper plot is calculated for narrow collimated kicks within 
$\theta=10^{\circ}$. The bottom plots shows the result for
isotropic kicks ($\theta=90^{\circ}).$
\label{mb_3d8}
}
\end{figure*}

Fig. ~\ref{t_cum_lgA}  
shows the integral distribution of the binary NS coalescence rate as a function of 
the time of coalescence (determined as the time interval from the formation 
of a double NS binary until its merging due to gravitational wave emission) 
for different models of NS formation (models A and B).
The total (over galactic lifetime) binary coalescence rate is a few times 
higher in Model B than in Model A (see also Postnov and Kuranov 2008), 
which is naturally related to decrease in the binary disruption probability
during formation of low-kick NSs in Model B. A strong dependence of the
NS binary coalescence rate on the kick-spin alignment effect 
characterized by the value of the angle $\theta$ can also bee seen. For small
$\theta$ (strong alignment) the mean time of NS binary coalescence increases, so
mainly old NSs coalesce. With increasing angle $\theta$ (the kick isotropisation) 
the fraction of young coalescing NSs with high magnetic field increases, since
for isotropic kicks there is a certain probability for the NS to get kick directed 
oppositely to the orbital motion at the moment of explosion. Correspondingly, the 
fraction of coalescing highly magnetized NSs is maximum for isotropic kicks
when $\theta=90^{\circ}$, see Fig.~\ref{mb_3d8}.

\begin{figure*}[ht]
\centering
\includegraphics[height=0.65\textwidth, width=1.0\textwidth]{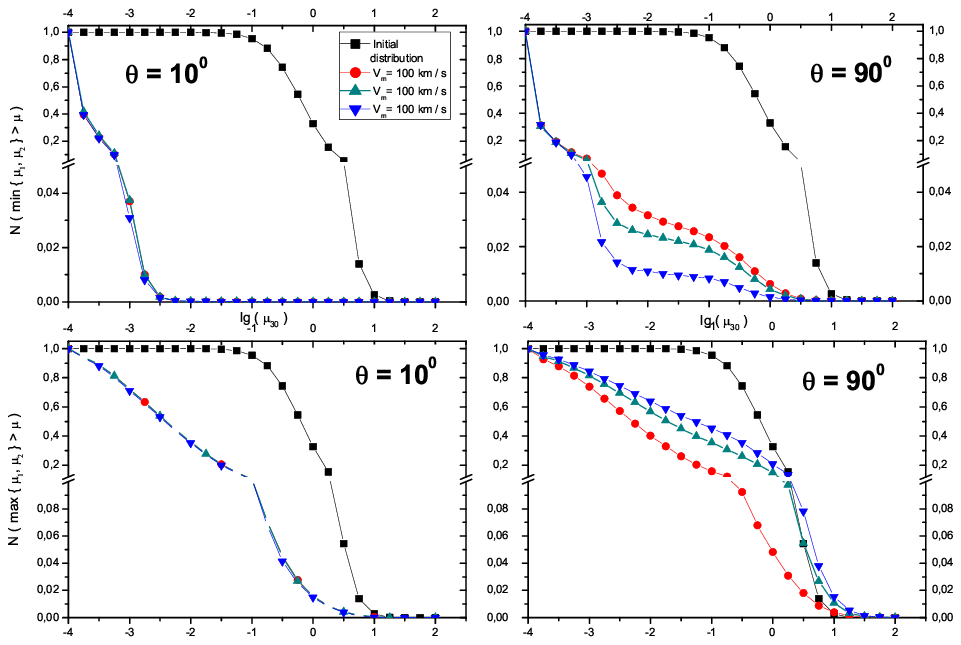}
\includegraphics[height=0.65\textwidth, width=1.0\textwidth]{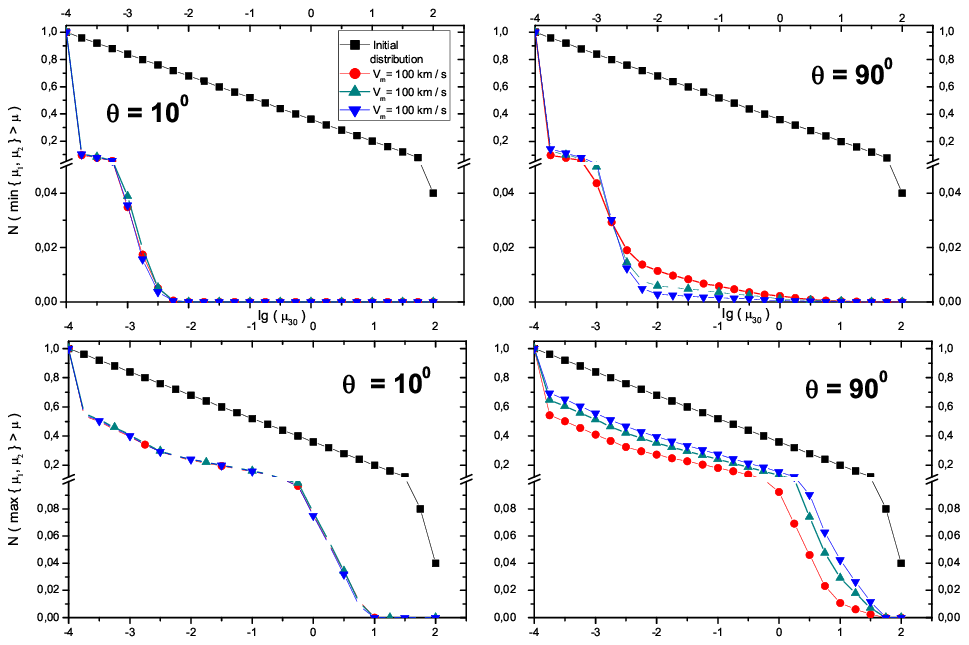}
\caption{The integral distribution function of the minimum (to the left) and
maximum (to the right) dipole moment of two binary NS components by the time of
coalescence. The upper and bottom plots are calculated for Model BG8 and BF8, 
respectively. The angle $\theta$ for natal kick-spin correlation is also shown.
\label{mb_min_max}
}
\end{figure*}

\begin{figure*}[ht]
\centering
\includegraphics[height=1.0\textwidth, width=1.0\textwidth]{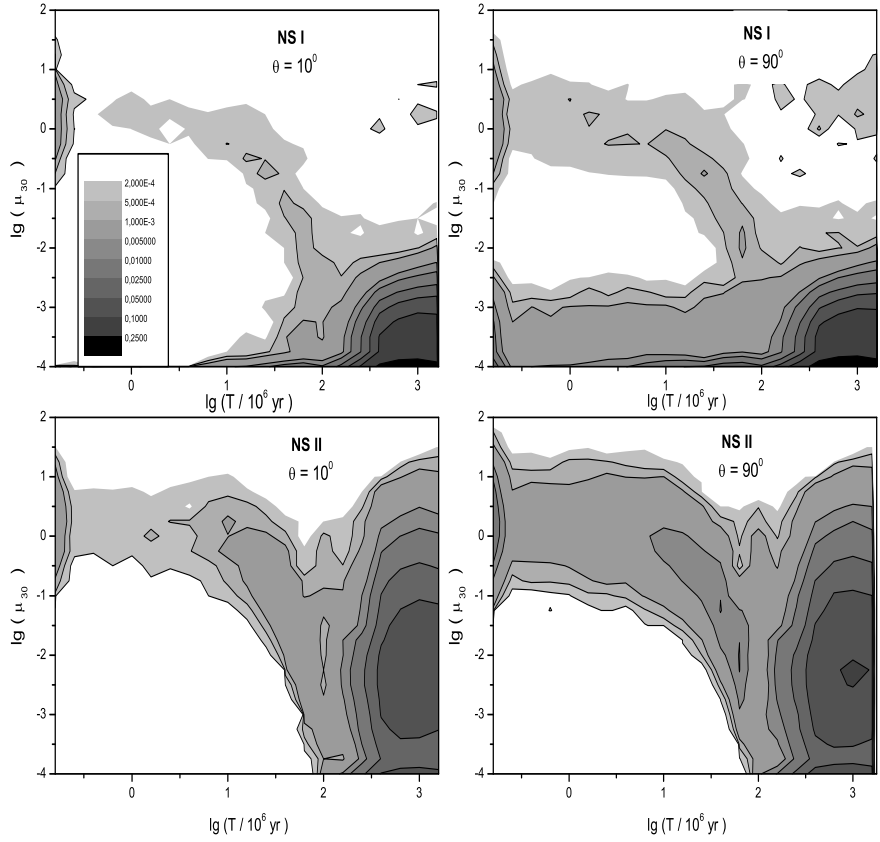}
\caption{The distribution function of the dipole magnetic moments of 
the poloidal magnetic field and the coalescence time in Model BG8. 
The upper and bottom plot shows the distribution for the first (older) NSI
and the second (younger) NSII, respectively. Left and right plots 
are for narrow-collimated ($\theta=10^{\circ}$) and isotropic  
($\theta=90^{\circ}$) natal kicks, respectively.}
\label{mbg_3B100}
\end{figure*}

\begin{figure*}[ht]
\centering
\includegraphics[height=1.2\textwidth, width=1.2\textwidth]{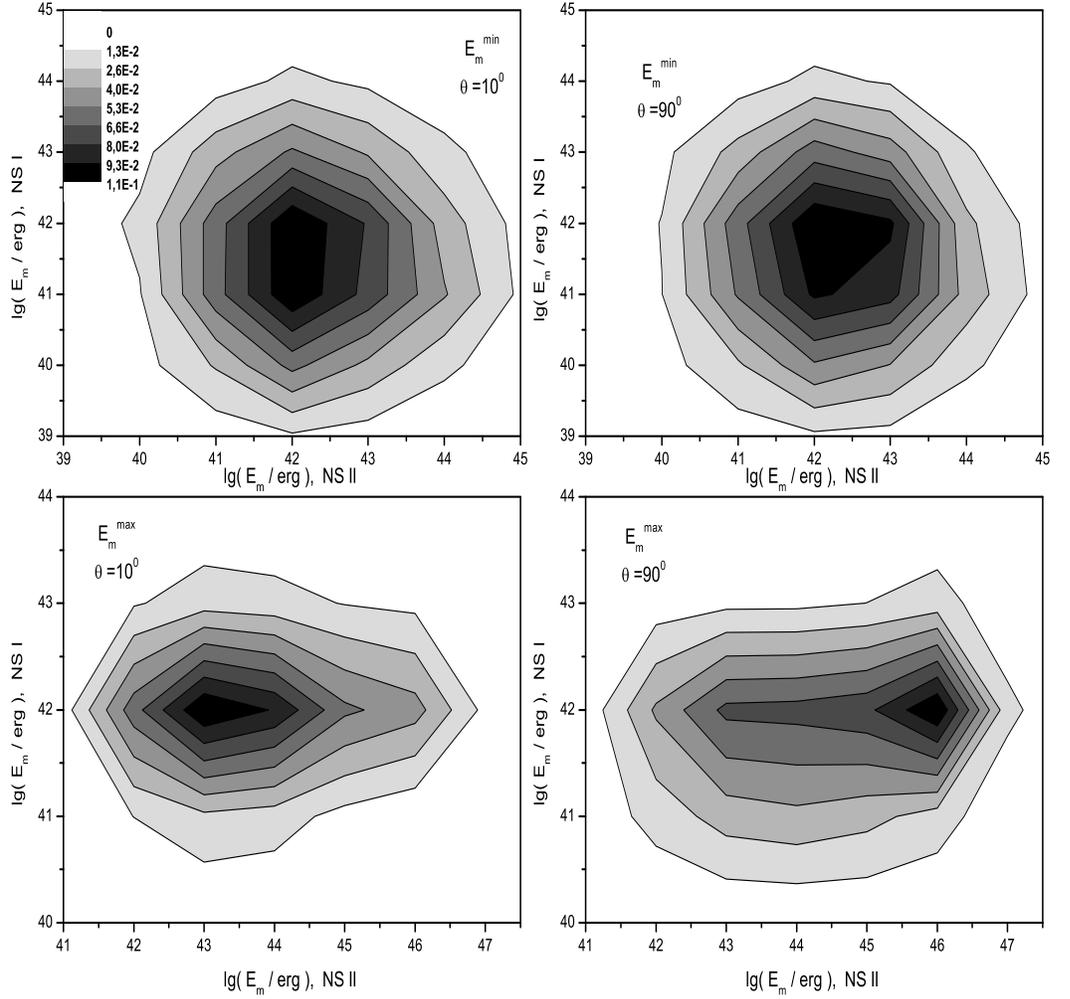}
\caption{The minimum ($E_m^{min}$, the upper panel) and 
maximum ($E_m^{max}$, the bottom panel) total magnetic field energy of 
coalescing binary NS components by the time of coalescence. Model BG8. }
\label{minmaxB}
\end{figure*}

Fig.~\ref{mb_3d8} shows the distribution of NS dipole mometa 
(in units $\mu_{30}=\mu/(10^{30}$~G~cm$^3)$) at the coalescence 
for the tight kick-spin alignment (upper panel) and the isotropic kick
(bottom panel) for a log-normal initial magnetic field distribution 
(Model BG8).  In Fig.~\ref{mb_min_max} an integral distribution of
the minimum (left plots) and maximum (right plots) magnetic moments
of the components of coalescing double NSs are presented for different models. 
Plots in upper and bottom panels are calculated for the initial log-normal (Model BG8) 
and  log-flat (Model BF8) field distributions, respectively. The two-dimensional
distribution of magnetic moments of both components at coalescence is plotted in
Fig.~\ref{mbg_3B100} for the log-normal initial distribution (model BF8).
The minimum $E_m^{min}$ and maximum $E_m^{max}$ values of the total magnetic field energy 
of the coalescing binary components (satisfied to the stability criterion 
(\ref{stability})) in Model BG8 is shown in Fig.~\ref{minmaxB}.

\Section{Discussion}

The calculated magnetic field distribution of coalescing NSs has several 
astrophysical implications.

\textbf{1) Electromagnetic precursors of short gamma-ray bursts. }

A colaescing NS binary that leads to a
powerful gamma-ray burst (Blinnikov et al. 1984, Eichler et al. 
1989, see Nakar 2007 for a review) can be preceded by an electromagnetic
burst (Lipunov and Panchenko 1996). Assuming the Goldreich-Julian current 
in the relativistic wind along open magnetic field lines and the light cylinder 
radius  $R_l=c/\Omega \gg a$, which are determined dy the rotation 
of the magnetized NS, we obtain the standard estimation of the total EM losses
(e.g. Beskin 2009) 
\begin{equation}
W_{em}\sim (\Omega R/c)^4 B^2 R^2 c \sim \Omega^4\mu^2/c^3
\end{equation}
(here $\Omega=2\pi/P$ is the spin frequency of the NS, $R$ is its radius,  
$\mu=BR^3$ is the NS dipole magnetic moment, $a$ is the radius of the 
circular orbit, $c$ is the speed of light). Assuming that the NS rotation 
is synchronized with its orbital relvolution at final stages before 
the coalescence, i.e.  $\Omega^2= 2GM/a^3$ (the 3-d Kepler law for two 
equal-mass NS in the circular orbit with semi-major axis $a$), 
we obtain  
$$
W_{em}\sim 3 \times 10^{39} (\hbox{erg/s}) \mu_{30}^2
(M/M_\odot)^2 (a/10^7\hbox{cm})^{-6}\,.
$$
The characteristic values of the magnetic field of the second (younger) NS
in the binary in the considered models is 
$\mu_{30}(NSII)\sim 10-10^{-4}$ (see Fig.~\ref{mb_min_max}), and we obtain
a total EM power emitted before the coalescence within the range from 
$\sim 10^{32}$ to $\sim 10^{42}$~erg/s, with the energy release strongly 
increasing with orbital decrease as $1/a^6$. 

In the relativistic particle wind 
flowing along open field lines plasma instability can develop to generate 
high-frequency radio emission, as in radio pulsars. The efficiency of 
conversion of the total power released into radio emission is fairly small in 
radio pulsars,  $\sim 10^{-5}-10^{-6}$, so even during last orbital revolutions before
the coalescence the expected power of a radio burst is hardly to be more than 
$\sim 10^{42}$~erg/s. In principle, this power is suffcient 
to explain properties of the unique extragalactic 5-ms radio burst
(Lorimer et al 2007). However, such a powerful radio burst would require
a significan dipole magnetic field of the NS at coalescence, 
$\mu_{30}\gg 1$. Fig.~\ref{mb_min_max} shows that for the most 
favorable models with narrow kick-spin alignment the fraction of such highly
magnetized NS is of order of a few per cents of the total 
number of coalescing NS binaries. This aggravates the low-event-rate problem 
for the explanation of the obsserved ms raio burst by the binary NS coalescence
model (Lorimer et al 2007, Popov and Postnov 2007). On the other hand, in this model 
there is no problem for the highly coherent radio emission to go out of the
generation region due to induced scattering (Lyubarsky 2008), which must be overcome 
in other models for this radio burst (e.g. Egorov and Postnov 2009).  
The fact that the observed ms radio burst was not associated with a gamma-ray burst
could be due to different beaming angles for the pre-outburst radio emission and
the subsequent gamma-ray burst. Low statistics (one event), however, makes the statistical
arguments very weak, so the possibility to explain this radio burst by mechanism
proposed by Lipunov and Panchenko (1996) still remains.  

\textbf{2) The effect of the magnetic field of coalescing NSs on the form of the 
gravitational wave signal}

Yuk Tun Li et al (2008) studied numerically the effect of a strong magnetic field of
the components of a coalescing NS binary on the form of the emitted gravitational wave (GW) signal 
at the stage preceding the merging of the components. In their numerical relativistic calculations, these authors used models of non-rotating NSs with a strong
($B\sim 10^{16}$~G) poloidal magnetic field. The difference in the GW signal form turned out
to be significant if the final collapse of the supramassive differentially rotating NS 
(formed in the merging) into a black hole is magnetically delayed. Our calculations show that 
even for the initial high (magnetar-like) magnetic fields, the vast majority of coalescing NSs 
has significantly (by orders of magnitude) smaller magnetic fields if the 
field decay is assumed (Fig.~\ref{mbg_3B100}). However, internal toroidal magnetic field
can be much higher than the poloidal component, so that the energy of the magnetic field 
at coalescence can be mostly determined by the toroidal component and be as high as 
$10^{47}$~ergs (and even more if one assumes the initial fields as high as  
 $\sim 10^{15}-10^{16}$~G). Although such a field can not be too significant
dynamically ($E_{m}/|E_{gr}|\sim 10^{-5}-10^{-6}$), it may serve as a seed field 
for the magnetic field generation in a merged differentially rotating object. In principle, 
following arguments by Spruit (2008), the energy of differential rotation of the pre-collapsed merged
object may entirely be transformed into the magnetic field energy, so that 
\begin{equation}
B^2R^3\sim (\Delta \Omega/\Omega)^2 E_{rot}\,,
\end{equation}
where $R$ is the radius of the object, the factor $\Delta \Omega/\Omega$ 
takes into account the differential rotation and $E_{rot}$ is the rotational energy. 
Assuming similar field generation processes to operate in a differentially rotating proto-NS
and the pre-collapsing post-merging object, the magnetic field can be of order to 
$B_{max, coal}\sim B_{max, NS}(E_{rot, coal}/E_{rot, NS})^{1/2}$. 
Limiting $E_{rot, NS}$ from above by the value $\sim 10^{51}$~ergs (Spruit 2008) and
assuming the virial value of $E_{rot, coal}$ of order of the binding energy of 
the NS $\sim 10^{53}$~ergs, we obtain the potentially possible energy of the poloidal 
magnetic field of pre-collapsing post-merging object about 
$\sim 10$ times higher than for a young NS. This energy can be dynamically significant 
for the subsequent evolution of this object and must be taken into
account in calculations of the expected form of the GW signal at post-merging
phases. 

\textbf{3) The luminosity function of short gamma-ray bursts}

Probably, the most intriguing implication of our calculations relates to 
short gamma-ray bursts. Presently, the merging of compact NS binaries 
is thought to be the principal formation channel of short hard gamma-ray bursts
(Blinnikov et al. 1984, Eichler et al. 1989, see Nakar 2007 for a recent review). 
The plausible physical mechanism of generation of hard electromagnetic gamma-radiation
during the NS binary coalescence is related to the formation of ultra-relativisitc MHD jet 
(Usov 1992, see recent numerical calculations by Komissarov et al. 2009). The NS magnetic
field can be an important initial parameter for such a jet formation. Indeed, 
the power of 
energy release in the MHD-mechanism is roughly $L\sim B^2\omega ^4 R^6$, where $\omega$ 
is the characteristic spin frequency, $R$ is the radius and $B$ is the magnetic field
strength. For different coalescing NS binaries parameters $\omega$ and $R$ are most likely 
similar, while the value of the magnetic field strength can be significantly different. Of course,
as we argued above, the magnetic field can strongly increase during the coalescence process, 
however immediately before the merging NS fields serve as the initial conditions.
Thus, \textit{there are physical grounds to expect that the distribution of luminositites of short gamma-ray bursts
$\Phi(L)$ will reflect the distribution of magnetic fields of coalescing NSs}:
$d \Phi(L)/dL \propto d N/d B$.  

We shall consider two different possibilities.

1) $\Phi(L)$ reflects distribution of the energy of the NS poloidal field 
 $E_p~B^2_p$ before the coalescence, 
$$
\Phi(L)\propto dN/d E_p\propto dN/d(B_p^2)\,.
$$ 
and 

2)  $\Phi(L)$ reflects the distribution of the total magnetic field energy 
(including the toroidal component) $E_m=E_p+E_t$. According to the stability criterion 
for the poloidal magnetic field (\ref{stability}), the maximum total energy 
of the magnetic field scales as $E_m\sim \sqrt{|E_{gr}|E_p}\propto B_p$, 
i.e. for a given value of the poloidal field strength 
$B_p$, the distribution of the maximum possible magnetic field energy 
in a stable configuration 
is proportional to that of the poloidal magnetic field strength, 
$$
\Phi(L)\propto dN/d E_m\propto dN/dB_p\,.
$$ 

\begin{figure*}[ht]
\centering
\includegraphics[height=1.1\textwidth, width=1.0\textwidth]{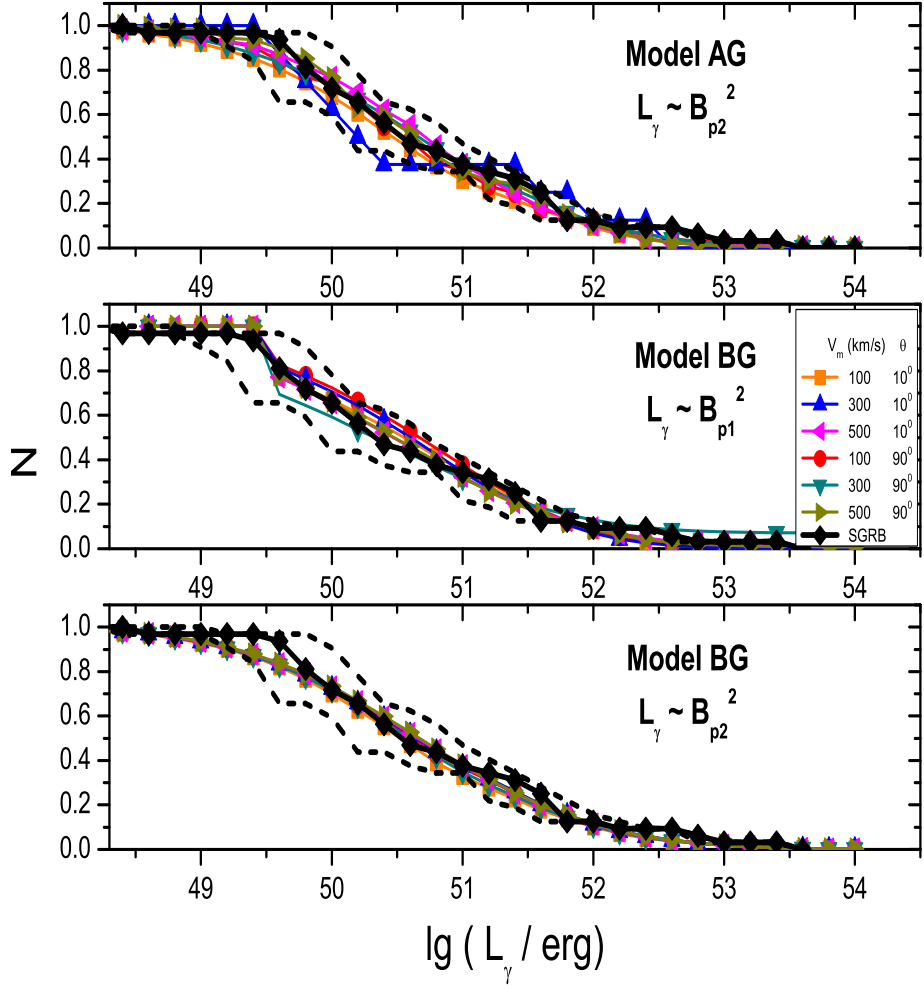}
\caption{The comparison of the hypothesis
$L_\gamma\propto B_p^2$ for the luminosity of short gamma-ray bursts 
for different NS formation parameters in models AG and BG. On the top and
bottom panel the comparison with the distribution of the poloidal field 
of the second (younger) NS in the binary is shown, while the plot on
the middle panel is calculated for the field distribution of the first
(older) NS.  The observed integral distribution function of short gamma-ray bursts
$N(>L)$ as computed from data presented in Table 1 of Kann et al. (2008) is
shown by the solid line connecting the filled diamonds. The dashed lines indicate
1-$\sigma$ errors of the gamma-rya burst luminosity  $L_\gamma$.}
\label{KSBp}
\end{figure*}

\begin{figure*}[ht]
\centering
\includegraphics[height=1.2\textwidth, width=1.0\textwidth]{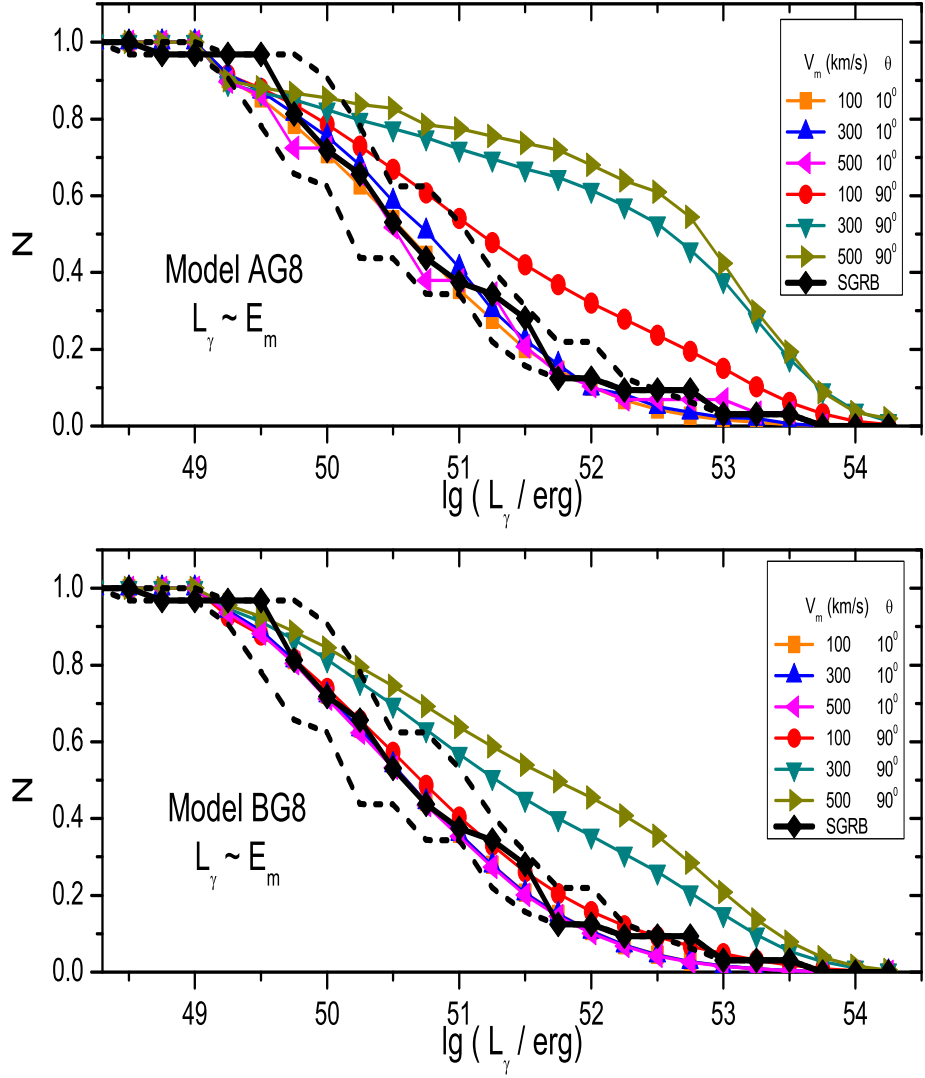}
\caption{The comparison of the hypothesis 
$L_\gamma\propto E_m$ for the luminosity of short gamma-ray bursts 
in models with exponential field decay AG8 (the upper panel) 
and BG8 (the bottom panel) for the initial log-normal distribution of 
the poloidal magnetic field component of NS.}
\label{KSE}
\end{figure*}

Our calculations of magnetic fields of coalescing NS binaries 
allows testing these hypotheses. To this aim, we shall use the luminosity 
function of short gamma-ray bursts calculated using observational data
for short gamma-ray bursts obtained 
by space observatories \textit{Swift, INTEGRAL, Fermi}
(Kann et al 2008). In Figs.~\ref{KSBp} - \ref{KSE} 
the integral luminosity funciton of short gamma-ray bursts
$N(>L)$ is shown by the solid black line connecting solid diamonds. 
The dashed lines show 1-$\sigma$ errors of the total GRB luminosity 
estimation $L_\gamma$ (assuming spherical symmetry of emission) 
taken from Table 1 of Kann et al. (2008). 
In Fig.~\ref{KSBp} we compare the hypothesis 
$L_\gamma\propto B_p^2$ for different initial NS parameters  
(the natal kick velcoity parameter $V_m$, 
spin-kick alignment angle $\theta$) in our Models AG and BG
(in which the NS magnetic field can decay only during accretion) with 
log-normal initial distribution of the poloidal magnetic field. 
The upper and lower panel of  Fig.~\ref{KSBp} show the comparison with 
the poloidal component of the second (younger) NS in the binary, the middle panel 
show the comparison with the field of the first (older) NS. 
It is seen that the observed short GRB luminosity distribution is 
in agreement with the distribution of magnetic fields of NS before the coalescence 
for any assumptions on the value and the direction of the natal NS kick velocity. 

In Fig.~\ref{KSE} we compare 
the observed luminosity function of short gamma-ray bursts with the distribution of 
the total energy of the magnetic field before the coalescence
(i.e. the hypothesis $L_\gamma\propto E_m\propto B_p$) for models with 
exponential field decay AG8 (the upper panel) and BG8 (the bottom panel), as well 
as for a log-flat initial field distribution. For these models, the agreement with 
observations is reached for all values of narrow-collimated kicks 
($\theta=10^0$), and cannot be reached at high values of 
the isotropic kicks   
($V_m=500$, 300 km/s and  $\theta=90^0$). The parameter of the Kolmogorov-Smirnov test
for different models is listed in Table 2.  

Note that in models with  a logarithmically flat initial magnetic field distribution 
(our models AF, AF8, BF, BF8) the observed luminosity function of short gamma-ray bursts
can not be described for any value and direction of the NS kick. 

\Section{Conclusions}

Using the population synthesis method (the Scenario Mashine code), 
we have calculated the expected distribution of magnetic fields of coalescing binary neutron
stars for various initial distributions and models of NS formation. We have 
taken into account both poloidal and toroidal components of the field
satisfying the field stability criterion. The most important 
parameters of NS formation include the form (single-mode, bimodal 
Maxellian distribuitons) and amplitude (30-500 km/s) of the natal kick velocity 
and its possible alignment with the NS spin. We have taken into account the 
NS magnetic field evolution (no decay, exponential decay until a bottom
value of $10^8$~G, accretion-induced decay). We discuss possible 
observational manifestations of the strong magnetic field of coalescing neutron stars
(the electromagnetic precursors) and the effect of the magnetic field on the form of
gravitational wave signal. 

We have shown that the observed luminosity function of short/hard gamma-ray bursts
is very well fitted by the magnetic field distribution of the coalescing neutron star binaries
(the Kolmogorov-Smirnov test of order of one) for different initial NS formation models if
the initial magneitc field distribution is taken in the log-normal form with parameters
inferred from the radio pulsar observations. The obtained agreement of the distribution of
magnetic fields of coalescing NS with the observed luminosity funciton of short
gamma-ray bursts lends further support to the pioneer idea by Blinnikov et al. (1984) that 
coalescing NS binaries can be progenitors of cosmic gamma-ray bursts.

\vskip\baselineskip

The work is supprted by RFBR grant 07-02-00961.

\newpage
\eject
\begin{table}
\caption{Models of the intital magnetic fields and natal kicks of NS}
\bigskip
\begin{center}
\begin{tabular}{lcccr}
\hline
      & Initial distribution  &  $B_p$ field decay       &    $V_\mathrm{m}$  Initial mass           \\
Model & of magnetic field         & time, yrs &  km s$^{-1}$ & of NS progenitor \\
\hline
 \bf{AG}&  $\log$--normal  & $\infty$  &100 ... 500 & $> 10M_\odot$ \\
\hline
       &         & & 30  &   $8M_\odot$--$11 M_\odot$\\
\bf{BG}& $\log$--normal  &$\infty$     &100...500 & $> 11M_\odot$\\
\hline
\bf{AG8}&  $\log$--normalе  & $10^8$  &100 ... 500 & $> 10M_\odot$ \\
\hline
       &         & & 30  &   $8M_\odot$--$11 M_\odot$\\
\bf{BG8}& $\log$--normal  & $10^8$   &100...500 & $> 11M_\odot$\\
\hline
\hline
\bf{AF}&  $\log$--flat   & $\infty$  &100 ... 500 & $> 10M_\odot$ \\
\hline
       &         & & 30  &   $8M_\odot$--$11 M_\odot$\\
\bf{BF}& $\log$--flat   &$\infty$     &100...500 & $> 11M_\odot$\\
\hline
\bf{AF8}&  $\log$--flat   & $10^8$  &100 ... 500 & $> 10M_\odot$ \\
\hline
       &         & & 30  &   $8M_\odot$--$11 M_\odot$\\
\bf{BF8}& $\log$--flat   & $10^8$   &100...500 & $> 11M_\odot$\\
\hline

\label{t_model}
\end{tabular}
\end{center}
\end{table}

\begin{table}
\caption{Kolmogorov-Smirnov test for the luminosity function 
of short gamma-ray bursts}
\bigskip
\begin{center}
\begin{tabular}{lccccc}
      & Tested   &  & &     \\
Model &hypothesis &$\theta$  & $V_\mathrm{m}$=100 km s$^{-1}$ & $V_\mathrm{m}$=300 km s$^{-1}$ &$V_\mathrm{m}$=500 km s$^{-1}$ \\
\hline
\hline
\bf{AG}&$L_{\gamma}\propto E_p$  &$10^\circ$ & \bf{0.904}  &\bf{0.995} & 0.525 \\
       &$L_{\gamma}\propto E_p$  &$90^\circ$ & \bf{0.991}  & \bf{0.999} & \bf{1.000} \\
\hline
\bf{BG}& $L_{\gamma}\propto E_p$ &$10^\circ$ &\bf{0.999}  & \bf{0.997} & \bf{1.000} \\
       & $L_{\gamma}\propto E_p$ &$90^\circ$ & \bf{0.893}  & \bf{0.947} &\bf{ 1.000} \\
\hline
\hline
\bf{AG8}& $L_{\gamma}\propto E_m$ &$10^\circ$ & \bf{0.997}  & \bf{0.999} & \bf{0.999} \\
       & $L_{\gamma}\propto E_m$ &$90^\circ$ & 0.469  & 2.6e-3 & 3.8e-4 \\
\hline
\bf{BG8}& $L_{\gamma}\propto E_m$ &$10^\circ$ & \bf{1.000}  & \bf{1.000} &\bf{ 1.000} \\
       & $L_{\gamma}\propto E_m$ &$90^\circ$ & \bf{0.999}  & 0.319 & 0.071 \\
\hline

\label{q_model}
\end{tabular}
\end{center}
\end{table}

\newpage
\eject

\end{document}